\documentclass[pre,twocolumn,nobalancelastpage,amsmath,amssymb,showpacs,
nofootinbib, superscriptaddress]{revtex4-1}

\usepackage{graphics}      
\usepackage{graphicx}      
\usepackage{longtable}     
\usepackage{url}           
\usepackage{bm}            
\usepackage{xcolor}
\usepackage{amsmath}
\usepackage{dcolumn}
\usepackage{mathtools}
\usepackage{hyperref}

\begin{document}
\title{Depletion Forces in Athermally  Sheared Mixtures of Frictionless Disks and Rods in Two Dimensions}
\author{Theodore A. Marschall}
\affiliation{Department of Physics and Astronomy, University of Rochester, Rochester, NY 14627}
\author{S. Teitel}
\affiliation{Department of Physics and Astronomy, University of Rochester, Rochester, NY 14627}
\date{\today}

\begin{abstract}
We carry out numerical simulations to study the behavior of an athermal mixture of frictionless circular disks and elongated rods in two dimensions, under three different types of global linear deformation at a finite strain rate: (i) simple shearing, (ii) pure shearing, and (iii) isotropic compression.  We find that the fluctuations induced by such deformations lead to depletion forces that cause rods to group in parallel oriented clusters for the cases of simple and pure shear, but not for isotropic compression.  For simple shearing, we find that as the fraction of rods increases, this clustering increases, leading to an increase in the average rate of rotation of the rods, and a decrease in the magnitude of their nematic ordering.  
\end{abstract}
\maketitle

\section{Introduction}

Entropic excluded volume forces are known to play a key role in systems of elongated, aspherical particles.  For hard rods in thermal equilibrium, Onsager \cite{Onsager} explained the isotropic to nematic phase transition by such effects.  As the particle packing increases, aligned particles have a smaller excluded volume.  While this  reduces the rotational entropy, it  causes an even greater increase in the translational entropy, causing the system to transition to an orientationally ordered phase. A similar effect, known as the depletion force, was proposed by Oosawa and Asakura \cite{Oosawa} to describe the effective attraction between large particles in a colloid of  smaller particles \cite{Adams}.  Depletion forces are observed not only in thermally equilibrated systems, but also in athermal but vibrated dry granular systems, in particular  mixtures of spheres and rods \cite{Galanis,Linan}.  Depletion forces are usually argued to be the basis for the ``Brazil nut effect" \cite{Duran,Sanders,Bose} in which, upon shaking, large particles rise to the top of a size-polydisperse mixture of athermal hard particles.

Here we ask whether depletion forces can arise in strictly athermal granular systems undergoing a uniform linear deformation.  When  local fluctuations in the granular system arise  solely from such global linear deformations, with no additional vibrations or mechanical agitation, can these fluctuations still  drive the entropic effects that give rise to depletion forces?   

Some of our previous work gives reason for doubt.  For size-bidisperse but shape-monodisperse systems of either only circular disks or  only elongated rods, where the ratio of big to small particle lengths is a modest 1.4, we have found the following.   Isotropic compression of athermal rods, unlike thermally equilibrated rods, gives no nematic ordering  as the packing increases \cite{MTCompress}.
Bidisperse circular disks \cite{Vagberg.PRE.2011} and bidisperse rods \cite{MTstructure} show no size segregation in steady-state simple shear;  indeed shearing tends to mix  different particle sizes when starting from initial configurations that are more ordered.  However, here we will give evidence that depletion forces do  arise when mixtures of elongated rods and circular disks in a suspending host medium are subjected to uniform, steady-state, simple or pure shearing.

\section{Model}
\label{Secmodel}

\begin{figure}
\centering
\includegraphics[width=2.in]{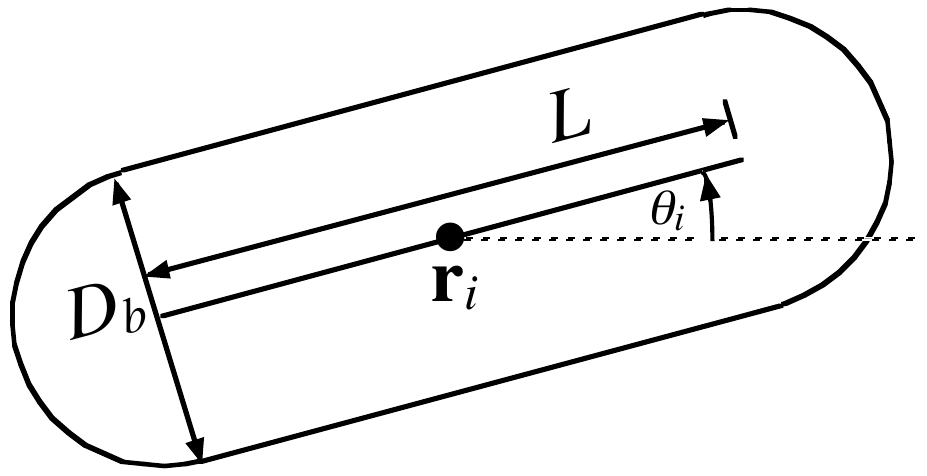}
\caption{Spherocylinder of axis length $L$, width $D_b$, and asphericity $\alpha=L/D_b$, centered at position $\mathbf{r}_i$ and oriented at angle $\theta_i$.
}
\label{sphero}
\end{figure}

Here we consider a two dimensional (2D)  athermal system  of $N$ total particles, of which a fraction $f$ are size-monodisperse rods, while the remaining  $1-f$ are size-bidisperse circular disks.  We take equal numbers of big and small disks with diameter ratio  $D_b/D_s=1.4$.  For  rods we use elongated 2D spherocylinders, composed of a rectangle of axis length $L$ caped by semi-circular endcaps of diameter $D_b$, as shown in Fig.~\ref{sphero}.  The asphericity of the spherocylinders is  $\alpha\equiv L/D_b=4$, giving a tip-to-tip length of $L+D_b=5D_b$.  We use $N=2048$ and consider systems with $N_\mathrm{rod}=1$, 64, 128, 256, and 512 rods, corresponding to fractions $f=0.00049$, 0.03125, 0.0625, 0.125, and 0.25.  A more geometric measure of the density of the rods is the ratio of the packing fraction of the rods $\phi_\mathrm{rod}$ to the total packing fraction $\phi$ of all particles.  With the packing fraction of the rods,
\begin{equation}
\phi_\mathrm{rod}=N_\mathrm{rod}\mathcal{A}_\mathrm{rod}/\mathcal{A}_\mathrm{tot},
\label{phirod}
\end{equation}
where $\mathcal{A}_\mathrm{rod}$ is the area of a rod and $\mathcal{A}_\mathrm{tot}$ is the total area of the system, and the packing fraction of the disks,
\begin{equation}
\phi_\mathrm{disk}=(N-N_\mathrm{rod})\frac{\pi}{2}(0.5^2+0.7^2)D_s^2/\mathcal{A}_\mathrm{tot},
\label{phidisk}
\end{equation}
the total packing fraction is,
\begin{equation}
\phi=\phi_\mathrm{rod}+\phi_\mathrm{disk}
\label{phitot}
\end{equation}
and so 
\begin{equation}
\dfrac{\phi_\mathrm{rod}}{\phi }=\dfrac{1}{1+\left(\frac{1-f}{f}\right)\left(\frac{0.74\pi D_s^2}{2\mathcal{A}_\mathrm{rod}}\right)}.
\end{equation}
Our cases for the above fractions $f$ then correspond to $\phi_\mathrm{rod}/\phi = 0.00392, 0.206, 0.350$, $0.536$, and 0.729. 

The forces on our particles are two fold: elastic contact forces when particles come into contact with each other, and dissipative drag forces with respect to a  suspending medium.
For the elastic contact interaction between particles we use a one sided harmonic potential as detailed in Ref.~\cite{MT1}.  A spherocylinder $i$ that is in contact with a  spherocylinder $j$  feels a force 
\begin{equation}
\mathbf{F}^\mathrm{el}_{ij}=(k_e/d_{ij})(1-r_{ij}/d_{ij})\mathbf{\hat n}_{ij}.
\label{eFel}
\end{equation}
Here $r_{ij}$ is the shortest distance between the axes of the two spherocylinders, $d_{ij}\equiv(D_i+D_j)/2$ is the average of the two spherocylinder widths, and 
$\mathbf{\hat n}_{ij}$ is the unit vector pointing normally inwards to spherocylinder $i$  at the point of contact with  spherocylinder $j$.  Two particles are in contact   whenever $r_{ij}<d_{ij}$.   The stiffness of the repulsion is $k_e$.  For contacts between a spherocylinder and a disk, or between two disks, we simply use the same Eq.~(\ref{eFel}), where  the disk is regarded as a spherocylinder with axis length $L=0$.

We model energy dissipation as a viscous drag between the  particles and a suspending background host medium \cite{MT1,MT2}.  If $\dot{\mathbf{r}}_i$ is the center of mass velocity of particle $i$, and $\dot\theta_i$ its angular velocity about the center of mass, then the local particle velocity at position $\mathbf{r}$ on  particle $i$ is 
\begin{equation}
\mathbf{v}_i(\mathbf{r})=\dot{\mathbf{r}}_i+\dot\theta_i\mathbf{\hat z}\times(\mathbf{r}-\mathbf{r}_i).  
\end{equation}
As a simplified model, we take the drag force to act everywhere over the area of the particle, with a force density proportional to the difference between the local velocity of the particle and the local velocity  of the host medium, $\mathbf{v}_\mathrm{host}(\mathbf{r})$,
\begin{equation}
\mathbf{f}^\mathrm{dis}_i(\mathbf{r})=-k_d[\mathbf{v}_i(\mathbf{r})-\mathbf{v}_\mathrm{host}(\mathbf{r})]. 
\label{eFdis}
\end{equation}
Integrating over the area of the particle then gives the total dissipative force on particle $i$, 
\begin{equation}
\mathbf{F}^\mathrm{dis}_i=\int_i d^2r\, \mathbf{f}_i^\mathrm{dis}(\mathbf{r}).  
\end{equation}
We are interested in the case of linear deformations, for which $\mathbf{v}_\mathrm{host}(\mathbf{r})=\dot{\boldsymbol{\Gamma}}\cdot \mathbf{r}$, where $\dot{\boldsymbol{\Gamma}}$ is a constant strain rate tensor.  In this case, integrating over $\delta\mathbf{r}_i=\mathbf{r}-\mathbf{r}_i$, one gets simply,
\begin{equation}
\mathbf{F}^\mathrm{dis}_i = -k_d\mathcal{A}_i [\dot{\mathbf{r}}_i-\mathbf{v}_\mathrm{host}(\mathbf{r}_i)],
\end{equation}
where $\mathcal{A}_i$ is the area of particle $i$.

The elastic and dissipative forces  give rise to elastic and dissipative torques on the particles.  The elastic torque on particle $i$ due to contact with particle $j$ is, 
\begin{equation}
\boldsymbol{\tau}_{ij}^\mathrm{el}=\mathbf{s}_{ij}\times\mathbf{F}^\mathrm{el}_{ij}, 
\end{equation}
where $\mathbf{s}_{ij}$ is the moment arm from the center of mass of $i$ to the point of contact with $j$.  Since $\mathbf{F}_{ij}^\mathrm{el}$ is always normal to the surface, for circular disks $\mathbf{s}_{ij}$ and $\mathbf{F}_{ij}^\mathrm{el}$ are always parallel and so the elastic torque always vanishes.  For the spherocylinders, however, $\mathbf{s}_{ij}$ and $\mathbf{F}_{ij}^\mathrm{el}$ are generally not parallel and so there can be a finite $\boldsymbol{\tau}_{ij}^\mathrm{el}$.

The dissipative torque is given by integrating the force density moment over the area of the particle,
\begin{equation}
\boldsymbol{\tau}_i^\mathrm{dis}=\int_id^2r\,(\mathbf{r}-\mathbf{r}_i)\times\mathbf{f}_i^\mathrm{dis}(\mathbf{r}).  
\label{etaudis}
\end{equation}
We will be interested in three different types of linear deformation at constant strain rate $\dot\gamma$: (i) simple shear with flow in the $\mathbf{\hat x}$ direction; (ii) pure shear, with compression along $\mathbf{\hat y}$ and expansion along $\mathbf{\hat x}$, both at the same rate; and (iii)  isotropic compression.  For these cases the host velocity is,
\begin{align}
\text{(i) } \quad&\mathbf{v}_\mathrm{host}(\mathbf{r})=\dot\gamma y\mathbf{\hat x}
\\
\text{(ii)\,} \quad&\mathbf{v}_\mathrm{host}(\mathbf{r})=\dot\gamma[x\mathbf{\hat x}-y\mathbf{\hat y}]/2
\label{epureshear}
\\
\text{(iii)} \quad&\mathbf{v}_\mathrm{host}(\mathbf{r})=-\dot\gamma\mathbf{r}.
\label{ecompress}
\end{align}
Using these in Eqs.~(\ref{eFdis}) and (\ref{etaudis}) then gives for the dissipative torque on particle $i$,
\begin{equation}
\boldsymbol{\tau}_i^\mathrm{dis}=-k_d\mathcal{A}_i I_i[\dot\theta_i+\dot\gamma f(\theta_i)]\mathbf{\hat z},
\label{etaudis}
\end{equation}
where \cite{MT1,MT2}
\begin{align}
\text{(i) } \quad& f(\theta) = [1-(\Delta I_i/I_i) \cos 2\theta]/2\\
\text{(ii)\,} \quad& f(\theta) = (\Delta I_i/I_i)[\sin2\theta]/2\\
\text{(iii)} \quad& f(\theta)=0.
\end{align}
Here $I_i$ is the sum of the two eigenvalues of the normalized moment of inertia tensor of particle $i$, while $\Delta I_i$ is the absolute value of their difference; in computing $I_i$ and $\Delta I_i$ we assume a uniform  mass density distributed over the area of the particle and normalize by the total mass of the particle \cite{MT1}.

Finally, we assume overdamped equations of motion for both the translational and rotational degrees of freedom.  With the total elastic force and torque on particle $i$ given by,
\begin{equation}
\mathbf{F}_i^\mathrm{el}={\sum_j}^\prime\mathbf{F}^\mathrm{el}_{ij},\quad
\boldsymbol{\tau}_i^\mathrm{el}={\sum_j}^\prime\boldsymbol{\tau}_{ij}^\mathrm{el},
\end{equation}
 where the sum is over all particles $j$ in contact with particle $i$, we have
\begin{equation}
\mathbf{F}_i^\mathrm{dis}+\mathbf{F}_i^\mathrm{el}=0,\quad
\boldsymbol{\tau}_i^\mathrm{dis}+\boldsymbol{\tau}_{i}^\mathrm{el}=0,
\end{equation}
which gives for the  translational and rotational equations of motion,
 \begin{align}
\dot{\mathbf{r}}_i&=\mathbf{v}_\mathrm{host}(\mathbf{r}_i)+\dfrac{\mathbf{F}_i^\mathrm{el}}{k_d\mathcal{A}_i}\label{er}\\
\dot\theta_i&=-\dot\gamma f(\theta_i) +\dfrac{\tau_i^\mathrm{el}}{k_d\mathcal{A}_i I_i}.\label{etheta}
\end{align}

Note that for the circular disks we have $\tau_i^\mathrm{el}=0$ and $\Delta I_i=0$.  Under simple shearing the disks will rotate with a constant angular velocity $\dot\theta_i = -\dot\gamma/2$.  Under pure shearing or isotropic compression, the disks do not rotate, and $\dot\theta_i=0$.

In contrast,  under simple shearing the rods  will in general rotate clockwise with a non-uniform angular velocity that varies according to the function $f(\theta)$ and the elastic torques $\tau_i^\mathrm{el}$ due to collisions.  Under pure shear the rods relax to orientations on average aligned with the minimal stress direction $\mathbf{\hat x}$, while under isotropic compression the rotation of rods is governed purely by the elastic torques $\tau_i^\mathrm{el}$.  Further details of the rotational motion of rods in our model can be found in Refs.~\cite{MTCompress,MT2}.

For our simulations we  take as the unit of length  $D_s=1$, the unit of energy $k_e=1$, and the unit of time $t_0=D_s^2 k_d\mathcal{A}_s/k_e=1$, where $\mathcal{A}_s$ is the area of a small disk.  For simplicity we choose the viscous drag $k_d$ to vary with particle size so that $k_d\mathcal{A}_i=1$ is the same for all particles.  We integrate using the Heun method with  step size $\Delta t/t_0=0.02$.  See \cite{MTCompress,MT1,MT2,SM} for
further details.   We start our simulations from an initial configuration in which particles are placed at random positions and rods have random orientations, however care is taken so  that no two rods have axes that intersect, as that would correspond to the unphysical situation of one rod penetrating through another.

\section{Results: Simple Shear}

\subsection{Depletion Forces}

\begin{figure}
\centering
\includegraphics[width=3.3in]{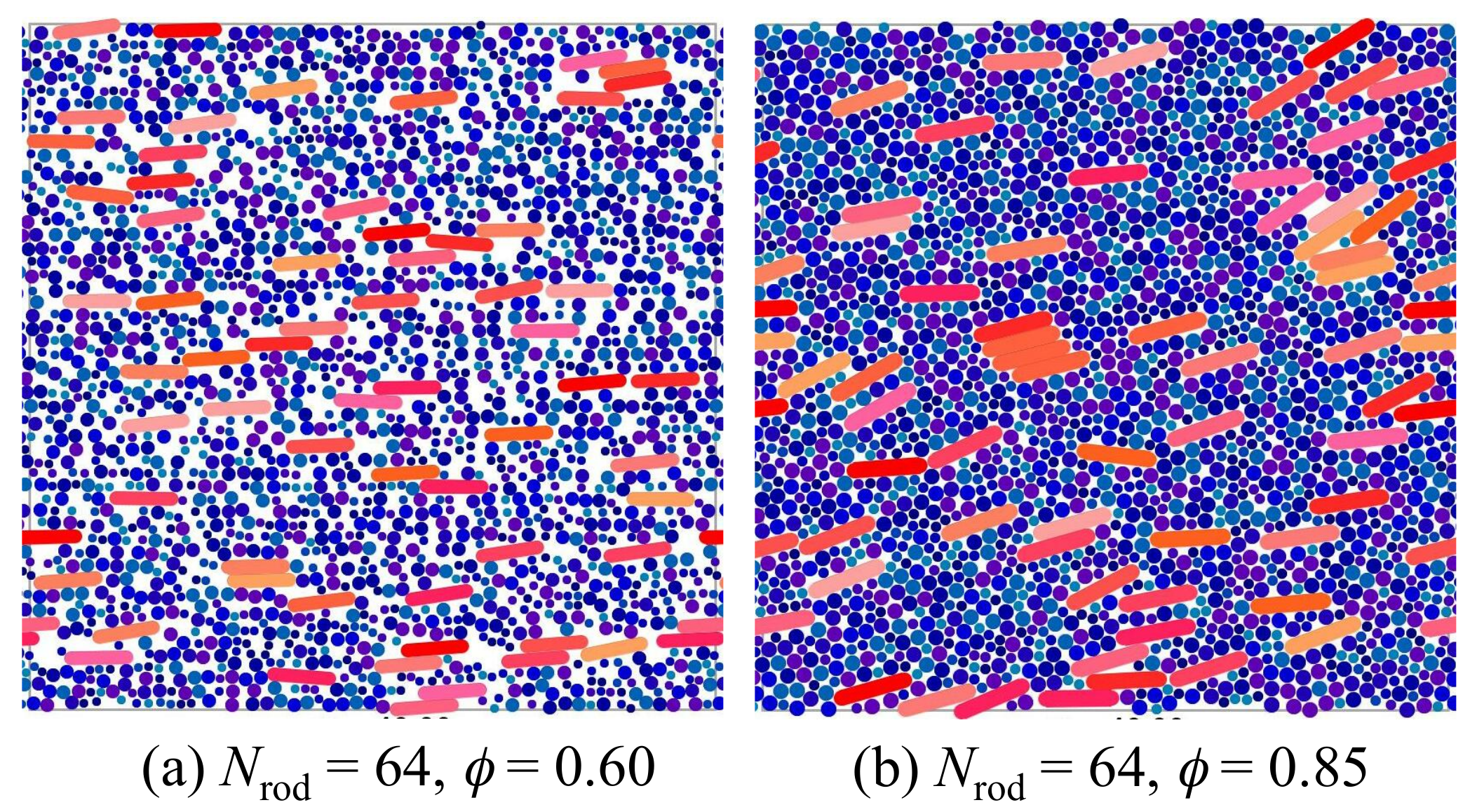}
\caption{Snapshots of  configurations in the sheared steady-state of $N_\mathrm{rod}=64$ spherocylinders in a sea of 1984 size bidisperse disks at packing (a) $\phi=0.60$ and (b) $\phi=0.85$.  Redish (light gray) hues are used for  spherocylinders, while bluish (dark gray) hues are  for  circular disks; in each case the different hues are used to help distinguish the individual particles, but have no further significance.  Systems are sheared at the rate $\dot\gamma=10^{-5}$.  Animations of these configurations are available in our Supplemental Material \cite{SM}.
}
\label{Nsp64}
\end{figure}

We first present our results for the case (i) of simple shearing, which is the main focus of this work.
We  shear  at the fixed rate $\dot\gamma=10^{-5}$, using Lees-Edwards boundary conditions to impose the shear strain \cite{LeesEdwards}.
In Fig.~\ref{Nsp64} we show snapshots of typical configurations in the sheared steady-state of a system with $N_\mathrm{rod}=64$ rods.
Fig.~\ref{Nsp64}(a) shows a configuration at the packing $\phi=0.60$, well below the jamming transition;  \ref{Nsp64}(b) shows a denser configuration at $\phi=0.85$, close to jamming.  In both cases one sees several pairs, and larger clusters, of rods in side-to-side contact, suggesting the action of depletion forces.  We also see examples where two parallel rods are separated by a single row of disks, as was previously observed in experiments on vibrated mixtures of rods and spheres \cite{Linan}.  Animations of these configurations, which show their evolution upon shearing from the initial random configuration, are included in our Supplemental Material \cite{SM}.  As seen in these animations, the clusters of rods in side-to-side contact are not static; they form, then separate under shearing, then  new clusters are formed.

To characterize the behavior of our system more quantitatively,  we first compute the stress tensor $\mathbf{p}$
 and the resulting pressure, $p=[p_{xx}+p_{yy}]/2$; the corresponding shear stress is $\sigma_{xy}=-p_{xy}$.  
 The stress tensor $\mathbf{p}$ is comprised of two pieces, one due to the elastic forces and one due to the dissipative forces \cite{MT1}.  The elastic part is
\begin{equation}
\mathbf{p}^\mathrm{el}=-\dfrac{1}{L_xL_y}\sum_{i=1}^N\boldsymbol{\Sigma}_i^\mathrm{el}, \quad
\boldsymbol{\Sigma}_i^\mathrm{el}={\sum_j}^\prime\mathbf{s}_{ij}\otimes\mathbf{F}_{ij}^\mathrm{el},
\end{equation}
where $\mathbf{s}_{ij}$ is the moment arm from the center of mass of particle $i$ to the point of contact with particle $j$, and the sum is over all particles $j$ in contact with $i$.  The dissipative part is
\begin{equation}
\mathbf{p}^\mathrm{dis}=-\dfrac{1}{L_xL_y}\sum_{i=1}^N\boldsymbol{\Sigma}_i^\mathrm{dis}, \quad
\boldsymbol{\Sigma}_i^\mathrm{dis}=\int_i d^2r\, (\mathbf{r}-\mathbf{r}_i)\otimes\mathbf{f}_i^\mathrm{dis}(\mathbf{r}),
\end{equation}
where $\mathbf{f}_i^\mathrm{dis}(\mathbf{r})$ is the dissipative force density of Eq.~(\ref{eFdis}) and the integral is over the area of the particle.
Further details may be found in Refs.~\cite{MT1,MT2}.  For most of our parameters, except at fairly low $\phi$, we find that the dissipative contribution $\mathbf{p}^\mathrm{dis}$ is negligible compared to the elastic contribution $\mathbf{p}^\mathrm{el}$.

\begin{figure}
\centering
\includegraphics[width=3.3in]{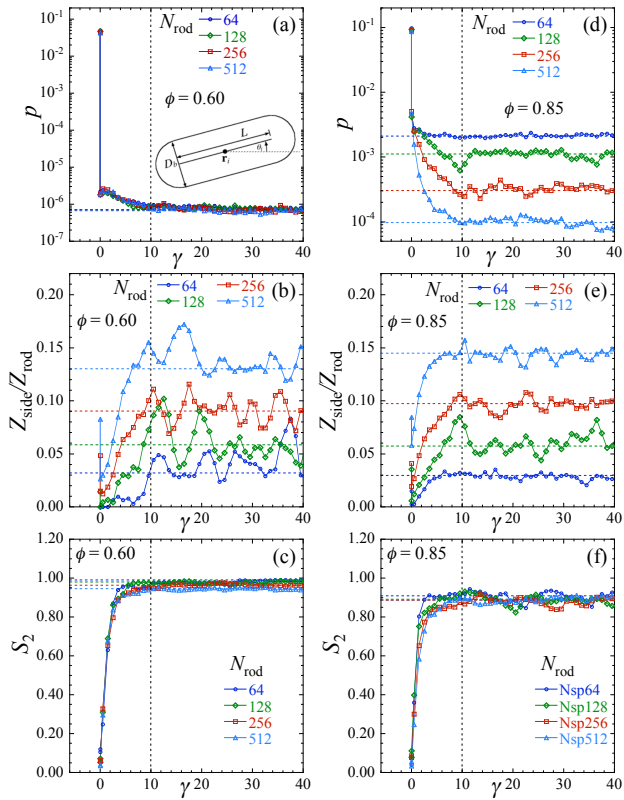}
\caption{For systems of $N_\mathrm{rod}=64$ to 512 spherocylindrical rods in a sea of size-bidisperse circular disks with $N=2048$ total particles, sheared at the strain rate $\dot\gamma=10^{-5}$ from initial configurations in which particles are placed at random and rods are placed with random orientations: (a) pressure $p$, (c) fraction of contacts on a spherocylinder that are side-to-side with another spherocylinder, $Z_\mathrm{side}/Z_\mathrm{rod}$,  and (e) magnitude of the nematic order parameter $S_2$, vs net strain $\gamma=\dot\gamma t$, for packing $\phi=0.60$.  Similarly, (b) pressure $p$, (d) fraction $Z_\mathrm{side}/Z_\mathrm{rod}$, and (f) $S_2$ vs $\gamma$, for $\phi=0.85$. 
Aside from the first two points, each data point represents an average of the instantaneous values  over a strain window  $\Delta\gamma=1$, so as to reduce fluctuations.  The horizontal dashed lines represent the average values in the steady-state, obtained by averaging over the last half of the run from $\gamma=50$ to 100.  The vertical dashed lines at $\gamma=10$ indicate roughly the strain over which the system relaxes to the steady-state.
In (b) the value of $N_\mathrm{rod}$ increases as the curves go from top to bottom; in (c) and (d) $N_\mathrm{rod}$ increases as the curves go from bottom to top.
}
\label{p-Z-S2-v-g} 
\end{figure}

In Fig.~\ref{p-Z-S2-v-g}(a) we plot the  pressure $p$ vs the net shear strain $\gamma=\dot\gamma t$, as the system is sheared at a packing $\phi=0.60$, well below jamming.  
We show results for systems with $N_\mathrm{rod}=64$, 128, 256, and 512 rods.  Because we start in a random initial configuration with many unphysically large particle overlaps, $p$ is initially large.  As we begin to shear, the system quickly relaxes these overlaps to small values,  pushing the particles away from each other.
The configurations obtained just after this initial quench are ones in which particles are  evenly distributed throughout the system, so as to avoid  large overlaps, but otherwise without any spatial correlations.  As the system is further sheared, the pressure continues to relax, but now more slowly.  Over a strain  of $\gamma\approx 10$ the particles evolve into configurations representative of the sheared steady-state, after which the pressure stays constant, aside from small fluctuations.

In Fig.~\ref{p-Z-S2-v-g}(b) we show the pressure $p$ at the larger packing $\phi=0.85$.  The behavior is qualitatively the same as in \ref{p-Z-S2-v-g}(a), except now one sees a much larger value of $p$, as well as a  large variation in the final steady-state values of $p$ as the number (and so the density) of the rods $N_\mathrm{rod}$ varies.  This is because of the proximity of the larger $\phi=0.85$ to the systems' jamming transition.  A system of only size-bidisperse disks has $\phi_J^{(0)}=0.8433$ \cite{OlssonTeitelPRE}, while a system of only size-monodisperse spherocylinders of $\alpha=4$ has $\phi_J^{(4)}\approx 0.92$ \cite{MTstructure}.  Our mixtures of disks and rods therefore have jamming transitions $\phi_J$ that vary between these two limits, with $\phi_J$ increasing as $N_\mathrm{rod}$ increases.  Since $p$ diverges as $\delta\phi = \phi_J-\phi$ vanishes, the mixtures at $\phi=0.85$ with smaller $N_\mathrm{rod}$ are closer to their system $\phi_J$ than are the mixtures with larger $N_\mathrm{rod}$, and so they have a larger $p$.  The variation of $\phi_J$ with $N_\mathrm{rod}$ is much less significant for the pressure at packings well below $\phi_J$, hence  a comparatively much smaller variation in the steady-state values of $p$ is seen Fig.~\ref{p-Z-S2-v-g}(a).

As a measure of the parallel clustering of rods we define the ratio $Z_\mathrm{side}/Z_\mathrm{rod}$, where $Z_\mathrm{rod}$ is the average number of contacts a rod has with any other particle, and $Z_\mathrm{side}$ is the average number of side-to-side contacts that a given rod has with other rods.  A side-to-side contact is when two rods make contact along their respective flat sides \cite{MTCompress}.  
In Figs.~\ref{p-Z-S2-v-g}(c) and \ref{p-Z-S2-v-g}(d) we plot, for $\phi=0.60$ and 0.85 respectively, the corresponding value of  
$Z_\mathrm{side}/Z_\mathrm{rod}$ vs $\gamma$.  Not surprisingly, we see that 
$Z_\mathrm{side}/Z_\mathrm{rod}$ increases as $N_\mathrm{rod}$ increases; the higher  the density of rods, the greater  the probability for there to be side-to-side contacts between them.  More  interesting, however, is the dependence of $Z_\mathrm{side}/Z_\mathrm{rod}$ on the shear strain $\gamma$ for fixed $N_\mathrm{rod}$.
As $\gamma$ increases, $Z_\mathrm{side}/Z_\mathrm{rod}$ first takes a sharp drop, from the value of the random initial configuration to a small value characteristic of the configuration in which the initial large overlaps  have relaxed, particles are more evenly spread throughout the system, but no correlations have yet been introduced by the shearing.  Then, as the shearing continues, $Z_\mathrm{side}/Z_\mathrm{rod}$ increases significantly, saturating to a constant value in the steady-state after a strain of roughly $\gamma\approx 10$, the same strain needed to relax the pressure to steady-state.  The strong correlation between the behavior of $p$ and $Z_\mathrm{side}/Z_\mathrm{rod}$ is simple to understand.  The clustering of rods with side-to-side contacts allows a more efficient packing of the system and thus a  decrease in the system pressure.  
Thus shearing acts to introduce a clustering among the rods, signaling the presence of depletion forces.    

Finally, we consider the orientational ordering of the rods.  It is well known that elongated particles in an athermal shear flow show nematic orientational ordering \cite{MT2,MKOT,Jeffery.RSPA.1922,Campbell,Guo1,Guo2,Borzsonyi1,Borzsonyi2,Wegner,Nagy,Trulsson}.  To quantify the orientational ordering we measure the magnitude  $S_2$ and orientation $\theta_2$ of the nematic order parameter.  In two dimensions, the magnitude $S_m$ and orientation $\theta_m$ of the  $m$-fold orientational order parameter can be written as \cite{Torquato}
\begin{align}
&S_m = \sqrt{\left[\dfrac{1}{N^\prime}\sum_{i}\cos m\theta_i\right]^2+\left[\dfrac{1}{N^\prime}\sum_{i}\sin m\theta_i\right]^2}\\
&\tan m\theta_m =\left[\dfrac{1}{N^\prime}\sum_{i}\sin m\theta_i\right]\Big/ \left[\dfrac{1}{N^\prime}\sum_{i}\cos m\theta_i\right].
\end{align}
For the instantaneous values of $S_m$ and $\theta_m$ in a given configuration, the above sums are over all the $N^\prime$ non-circular particles in that configuration.  For the ensemble average of $S_m$ and $\theta_m$, the terms $[\dots]$ in the above should be taken as averages over all configurations in the ensemble.  Here we are interested in the nematic orientational order, $m=2$.

In Figs.~\ref{p-Z-S2-v-g}(e) and \ref{p-Z-S2-v-g}(f) we plot, for $\phi=0.60$ and 0.85 respectively, the magnitude of nematic ordering $S_2$ vs $\gamma$.
We see that, similar to the behavior of $p$, $S_2$ rises rapidly from the value $S_2\approx 0$ of the random initial state, starts to plateau, but only reaches its steady-state value after the strain $\gamma\approx 10$.  One might think that the clustering of rods, as measured by the increased values of $Z_\mathrm{side}/Z_\mathrm{rod}$, is simply a consequence of the orientational ordering of the rods as the system is sheared.

Comparing Figs.~\ref{p-Z-S2-v-g}(c) and \ref{p-Z-S2-v-g}(d) with Figs.~\ref{p-Z-S2-v-g}(e) and \ref{p-Z-S2-v-g}(f), one might tend to think that it is  the orientational ordering of the rods in the shear flow that is the mechanism leading to the increase in side-to-side contacts between the rods, as the system is strained.  As the rods align orientations, it seems reasonable to think that side-to-side contacts become more frequent; hence the increase in $Z_\mathrm{side}/Z_\mathrm{rod}$ as we go from the initial disordered configuration with randomly oriented rods (and so $S_2=0$) to the steady-state configurations with aligned rods (and so $S_2 \approx 1$).  To show that this is not so, we have also considered shearing from initial configurations constructed as follows:  rods are placed uniformly throughout the system with orientations $\theta_i$ sampling the distribution $\mathcal{P}(\theta)$ found in the sheared steady-state; disks are placed at random.  Thus, in such initial configurations, the nematic ordering $S_2$ is  the same as found in steady-state, but there are few side-to-side contacts between rods.

\begin{figure}
\centering
\includegraphics[width=3.3in]{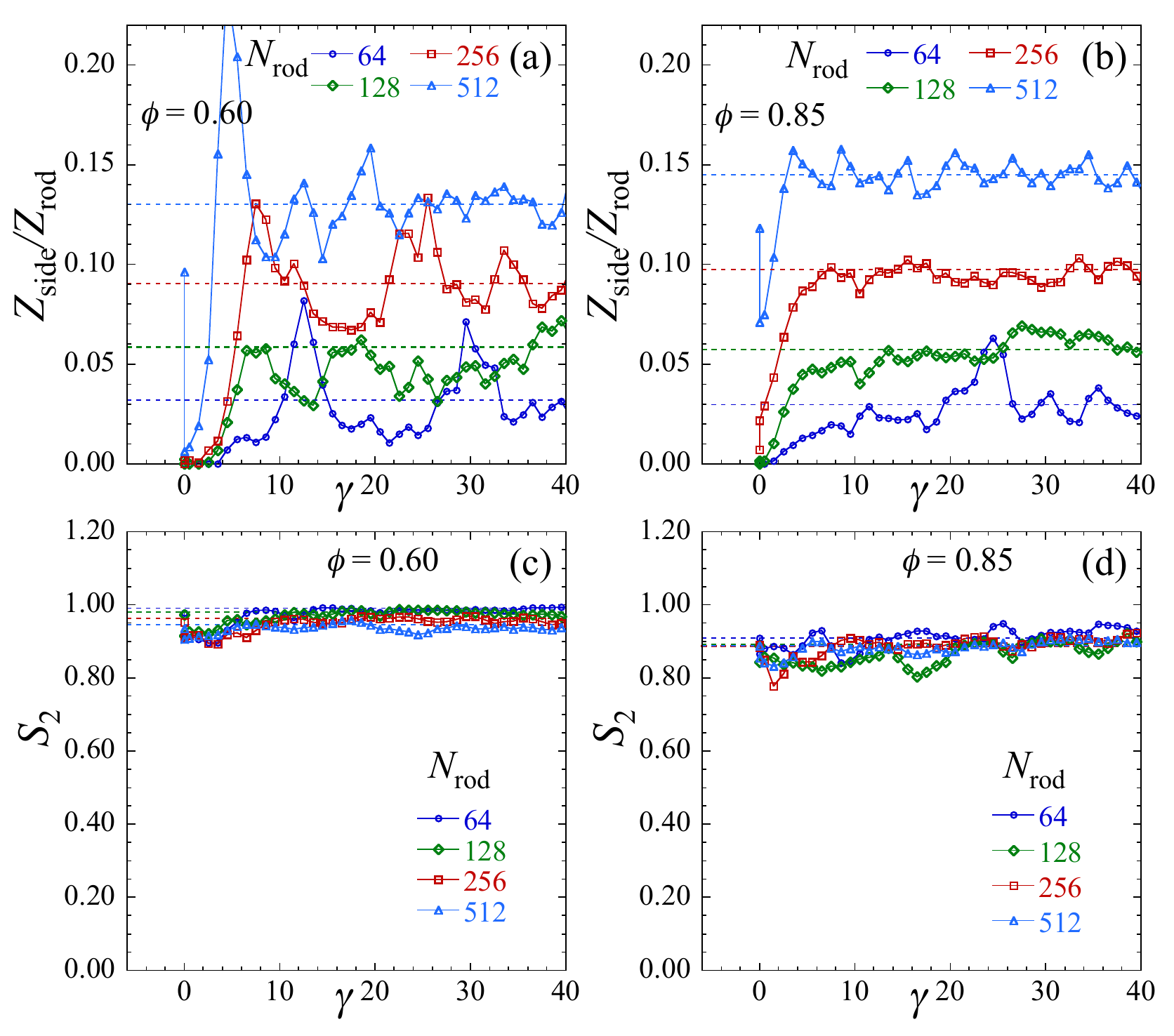}
\caption{For sheared systems from initial configurations in which rods are placed uniformly with orientations sampled from the distribution $\mathcal{P}(\theta)$ found in steady-state, and disks are placed randomly:
(a) fraction of contacts on a spherocylinder that are side-to-side with another spherocylinder, $Z_\mathrm{side}/Z_\mathrm{rod}$,  and (c) magnitude of the nematic order parameter $S_2$, vs net strain $\gamma=\dot\gamma t$, for packing $\phi=0.60$.  Similarly, (b)  fraction $Z_\mathrm{side}/Z_\mathrm{rod}$, and (d) $S_2$ vs $\gamma$, for $\phi=0.85$. 
Aside from the first two points, each data point represents an average of the instantaneous values  over a strain window  $\Delta\gamma=1$, so as to reduce fluctuations.  The horizontal dashed lines represent the average values in the steady-state.
Results are shown for $N_\mathrm{rod}=64$ to 512 spherocylindrical rods in a system with $N=2048$ total particles, sheared at the strain rate $\dot\gamma=10^{-5}$.   In (a) and (b) the value of $N_\mathrm{rod}$ increases as the curves go from bottom to top.
}
\label{p-Z-S2-v-g-2} 
\end{figure}

In Fig.~\ref{p-Z-S2-v-g-2} we show the resulting behavior of  $Z_\mathrm{side}/Z_\mathrm{rod}$ and $S_2$ as the system is strained from such an initial configuration.  We show results at the packings $\phi=0.60$ and 0.85.  We see in Figs.~\ref{p-Z-S2-v-g-2}(c) and \ref{p-Z-S2-v-g-2}(d) that $S_2$ remains relatively constant as the system is sheared.  In Figs.~\ref{p-Z-S2-v-g-2}(a) and \ref{p-Z-S2-v-g-2}(b), however, we see that $Z_\mathrm{side}/Z_\mathrm{rod}$ behaves similarly to what is seen in Fig.~\ref{p-Z-S2-v-g}.  As the system is strained, we find that $Z_\mathrm{side}/Z_\mathrm{rod}$ rises from a small value, after the quenching of overlaps in the initial configuration, to the larger value characteristic of the same steady-state found in Fig.~\ref{p-Z-S2-v-g}.   Thus an increase in the number of  side-to-side rod contacts is found even when the rods start from an orientationally ordered, but spatially uniform, initial configuration.  Orientation ordering is therefore not the mechanism for the increase in side-to-side contacts.

\begin{figure}
\centering
\includegraphics[width=3.3in]{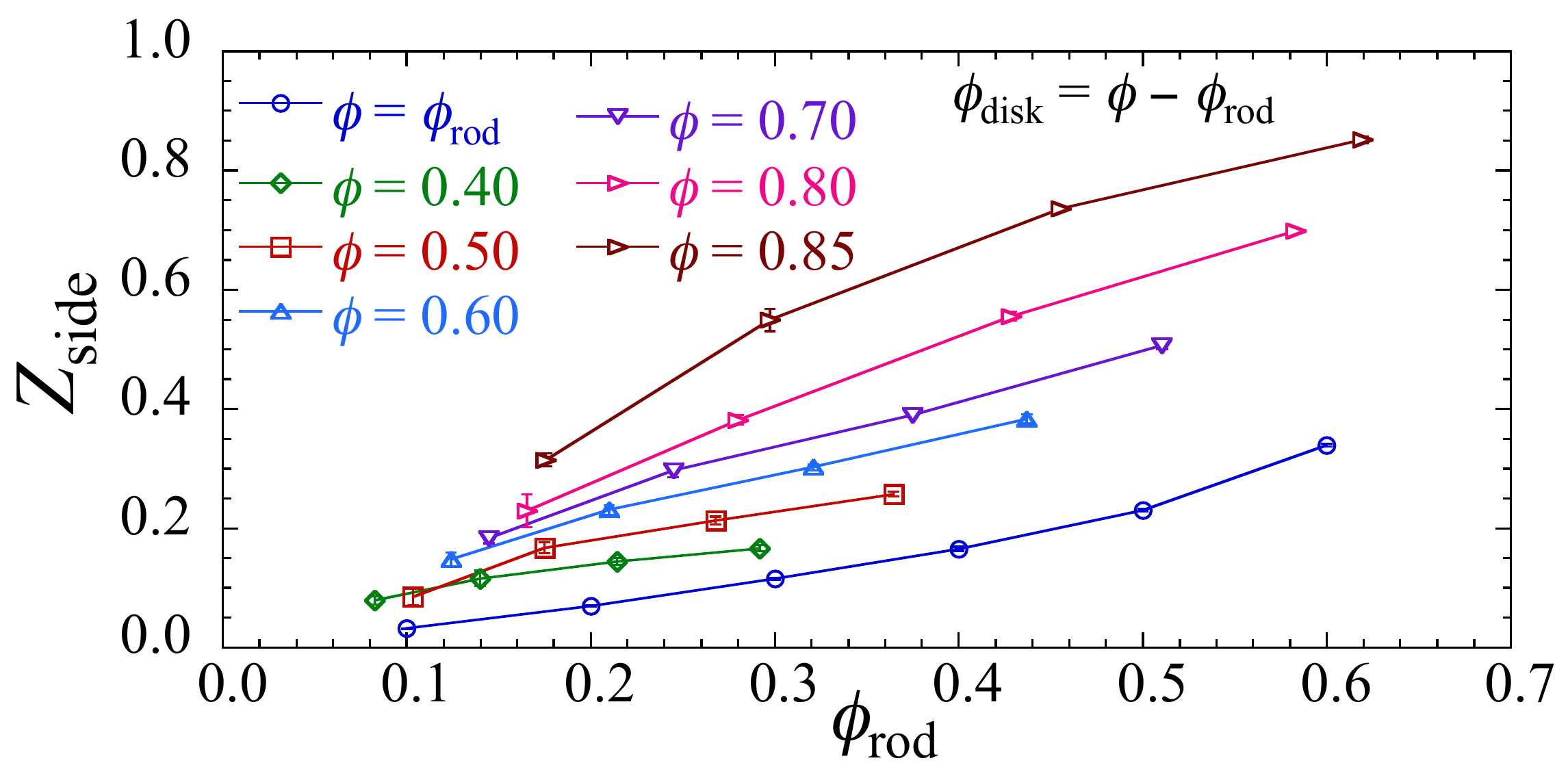}
\caption{Average number of side-to-side contacts $Z_\mathrm{side}$ that a given rod has with other rods in simple sheared steady-state at $\dot\gamma=10^{-5}$, vs packing fraction of rods $\phi_\mathrm{rod}$, for systems of different fixed total packing $\phi$.  Also shown is a system of only rods, $\phi_\mathrm{rod}=\phi$.  
As the curves go from bottom to top, $\phi$ increases as indicated.
We see that at fixed $\phi_\mathrm{rod}$, $Z_\mathrm{side}$ increases as $\phi$ increases, and hence as the density of disks, $\phi_\mathrm{disk}=\phi-\phi_\mathrm{rod}$ increases.
}
\label{Zside} 
\end{figure}

As another means of understanding the mechanism for the formation of side-to-side contacts of rods, we consider the behavior of our system as a function of the packing fraction of only the rods, $\phi_\mathrm{rod}$ of Eq.~(\ref{phirod}), rather than the total packing fraction of rods and disk, $\phi$ of Eq.~(\ref{phitot}).  We consider here the average number of side-to-side contacts $Z_\mathrm{side}$ that a given rod has with other rods, after the system has been strained sufficiently to reach the steady-state.  Considering a system at a fixed total packing $\phi$, varying the number of rods $N_\mathrm{rod}$ in that system is equivalent to varying $\phi_\mathrm{rod}$.  In this way, in Fig.~\ref{Zside} we plot $Z_\mathrm{side}$ vs $\phi_\mathrm{rod}$ for systems of different total packing $\phi$.  For comparison we also show $Z_\mathrm{side}$ for a system of only rods, i.e., $\phi_\mathrm{rod}=\phi$, at comparable rod packing densities.  Sitting at a fixed value of $\phi_\mathrm{rod}$ as the total $\phi$ increases, the different curves in Fig.~\ref{Zside} represent systems in which the packing $\phi_\mathrm{disk}$ of disks is increasing.  We see clearly that as the packing fraction of disks increases at fixed $\phi_\mathrm{rod}$, the number $Z_\mathrm{side}$ of side-to-side contacts between rods increases.  Thus it is the presence of the disks that facilitates the side-to-side contacts between the rods, supporting our claim that depletion forces lead to an effective attraction between the rods.

\subsection{Rheology}

\begin{figure}
\centering
\includegraphics[width=3.3in]{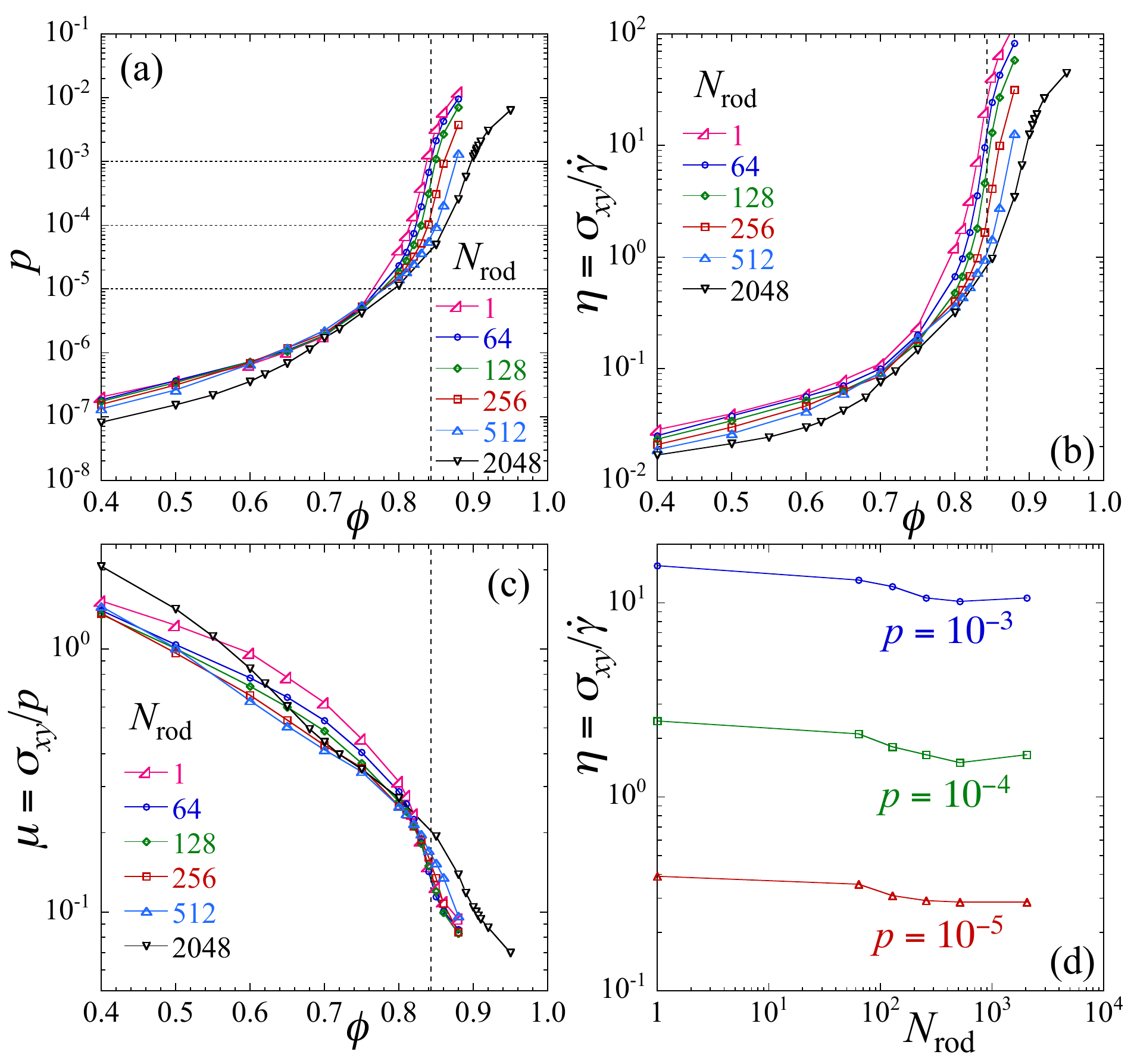}
\caption{
For $N_\mathrm{rod}=1$, $64$, 128, 256, 512, and 2048 spherocylindrical rods in  a system with $N=2048$ total particles: steady-state values of (a) pressure $p$, (b) shear viscosity $\eta=\sigma_{xy}/\dot\gamma$, and (c) macroscopic friction $\mu=\sigma_{xy}/p$ vs packing $\phi$; (d) $\eta$ vs $N_\mathrm{rod}$ at the three different values of constant pressure $p$ indicated by the horizontal dotted lines in (a).  For the case, $N_\mathrm{rod}=2048$, where all particles are rods, we use a size-bidisperse distribution of rods.  The vertical dashed lines indicate the jamming transition of size-bidisperse disks, $\phi_J^{(0)}=0.8433$.  Error bars are smaller than the symbol size.  In (a) and (b), for $\phi>\phi_J^{(0)}$, the value of $N_\mathrm{rod}$ increases as the curves go from top to bottom.
}
\label{p-eta-mu}
\end{figure}

It is now interesting to  examine the effect  that adding rods to a packing of disks has on the rheology of the system.  In Figs.~\ref{p-eta-mu}(a) and \ref{p-eta-mu}(b) we plot the steady-state pressure $p$ and shear viscosity $\eta=\sigma_{xy}/\dot\gamma$ vs packing $\phi$, for a fixed strain rate $\dot\gamma=10^{-5}$.  We show results for $N_\mathrm{rod}=1$, 64, 128, 256, and 512.  For comparison, we also show results for a system composed entirely of $N=2048$ spherocylinders; in this case we take a size-bidisperse distribution to avoid spatial ordering.  As  $\phi$ increases, the dependence on $N_\mathrm{rod}$ noticeably increases.  This is  due to the dependence of the jamming $\phi_J$ of the mixture on the density of rods, as discussed earlier in connection with Fig.~\ref{p-Z-S2-v-g}(b); we expect that $\phi_J$ must vary from $\phi_J^{(0)}=0.8433$ at a vanishingly low density of rods, to $\phi_J^{(4)}\approx 0.92$ as the system becomes mostly rods.  Thus, at a fixed large packing $\phi\gtrsim \phi_J^{(0)}$, we see that the shear viscosity $\eta$ decreases as more rods are added to the system.  In Fig.~\ref{p-eta-mu}(c) we show the macroscopic friction $\mu=\sigma_{xy}/p$ vs $\phi$.  In contrast to $\eta$, for fixed $\phi\gtrsim\phi_J^{(0)}$ we find that $\mu$ generally increases as $N_\mathrm{rod}$ increases.  
In experiments, one often creates packings under the condition of constant pressure rather than constant volume.  In Fig.~\ref{p-eta-mu}(d) we therefore show the shear viscosity $\eta$ vs $N_\mathrm{rod}$ at  three different fixed values of pressure $p$ that put the system close to jamming (see horizontal dotted lines in Fig.~\ref{p-eta-mu}(a)).  In each case  $\eta$ decreases slightly as $N_\mathrm{rod}$ increases. Note, for the situation in which both $p$ and $\dot\gamma$ are held constant, $\eta=\sigma_{xy}/\dot\gamma$ and $\mu=\sigma_{xy}/p$ are proportional.

\begin{figure}[h!]
\centering
\includegraphics[width=3.3in]{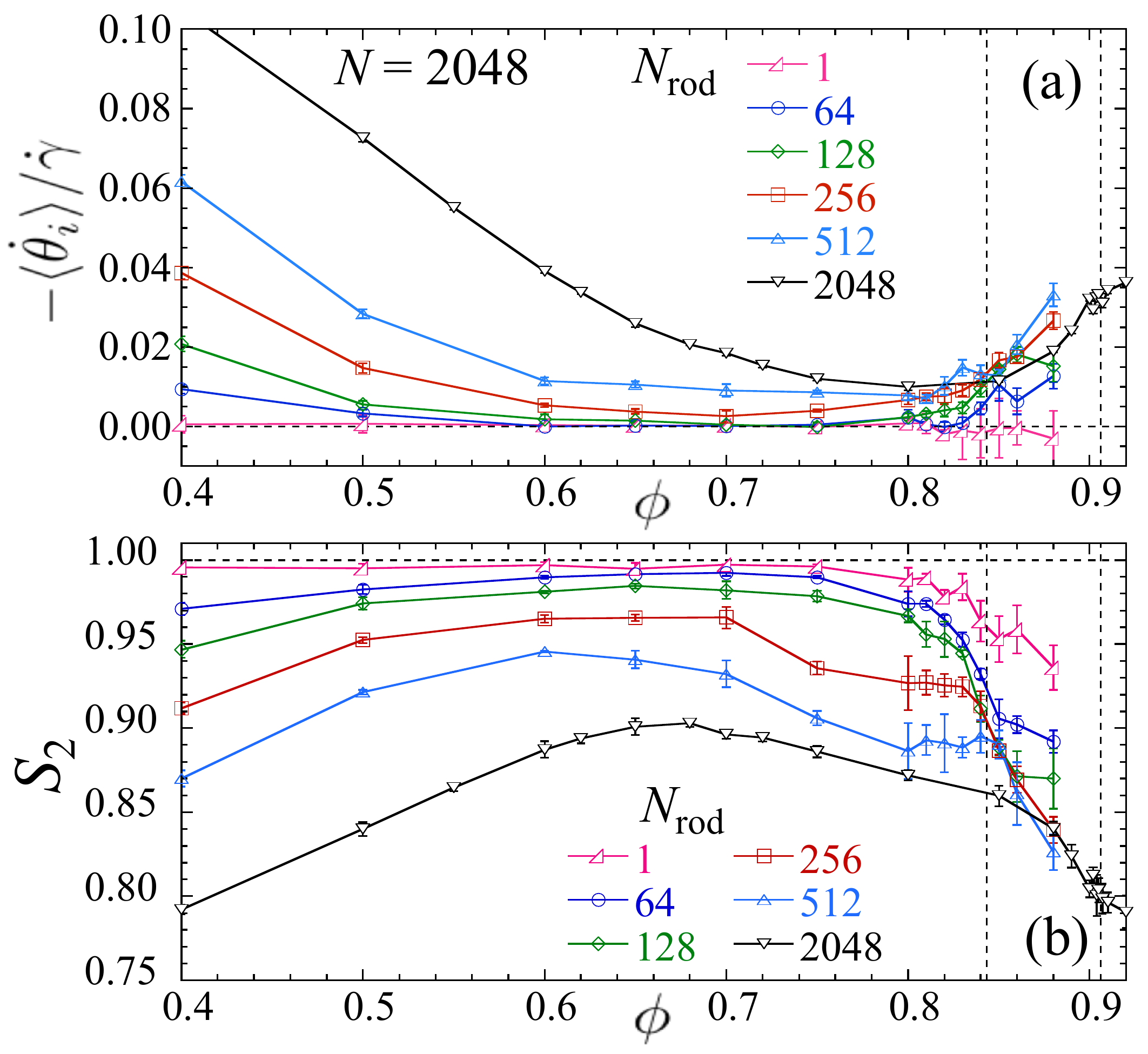}
\caption{(a) Average angular velocity $-\langle\dot\theta_i\rangle/\dot\gamma$, (b) magnitude of the nematic order parameter $S_2$, (c) orientation of the nematic order parameter $\theta_2$, and (d) contact ratio $Z_\mathrm{side}/Z_\mathrm{rod}$ vs packing $\phi$, for mixtures of size-bidisperse circular disks and size-monodisperse elogngated rods (i.e., sphererocylinders of $\alpha=4$).  The system has $N=2048$ total particles, and $N_\mathrm{rod}=1$, 64, 128, 256, 512 rods.
For comparison, results are also shown for a system of only $N=2048$ size-bidisperse spherocylinders.  The strain rate is $\dot\gamma=10^{-5}$.  The vertical dashed lines indicate the jamming transition of size-bidisperse disks, $\phi_J^{(0)}=0.8433$.
In (a) and (d) the value of $N_\mathrm{rod}$ increases as the curves go from bottom to top; while in (b), and for $\phi>\phi_J^{(0)}$ in (c), $N_\mathrm{rod}$ increases as the curves go from top to bottom.
}
\label{rods-disks} 
\end{figure}

Finally we examine  the rotational motion and orientational ordering of the rods in the simple shear flow.  As discussed in Sec.~\ref{Secmodel}, rods will experience torques from the elastic and dissipative forces that act on them, and the dissipative torque in particular will depend on the orientation of the rod, as given by Eq.~(\ref{etaudis}).  Thus rods will rotate non-uniformly, and exhibit a finite nematic orientational ordering  \cite{MT2,MKOT,Jeffery.RSPA.1922,Campbell,Guo1,Guo2, Borzsonyi1,Borzsonyi2,Wegner,Nagy,Trulsson}.  So it is interesting to see how such behavior is modified when the rods are immersed in a background sea of disks.
In Figs. ~\ref{rods-disks}(a) and \ref{rods-disks}(b) we plot the  steady-state average angular velocity scaled by the strain rate $-\langle\dot\theta_i\rangle/\dot\gamma$ and the steady-state magnitude $S_2$  of the ensemble averaged nematic order parameter vs the total packing fraction $\phi$, for the   different values of $N_\mathrm{rod}=1$ to  $512$.  In computing these quantities, we average only over the $N_\mathrm{rod}=fN$ rods, since the circular disks experience no collisional elastic torques and thus they rotate uniformly and do not order.  For comparison, we  show the same quantities for a  system of only $N=2048$ size-bidisperse, $\alpha=4$, spherocylinders.  

We see in Figs.~\ref{rods-disks}(a) and \ref{rods-disks}(b) that the behavior of the mixture of rods and disks is qualitatively similar to that of only rods \cite{MT2}.
The angular velocity $-\langle \dot\theta_i\rangle/\dot\gamma$ is non-monotonic, decreasing to a minimum and then increasing as $\phi$ increases.  The magnitude of the nematic order parameter $S_2$ is similarly non-monotonic, increasing to a maximum and then decreasing as $\phi$ increases.  
For the entire range of $\phi$ we see that as  $N_\mathrm{rod}$ decreases, $-\langle\dot\theta_i\rangle/\dot\gamma$ decreases, while $S_2$ increases; the fewer the rods, the more slowly they rotate and the more orientationally ordered they are.  
For the case of only a single rod, $N_\mathrm{rod}=1$, we see that  $-\langle\dot\theta_i\rangle/\dot\gamma\approx 0$ within the estimated errors and $S_2$ is close to unity.  This indicates that, for the range of $\phi$ shown, the angular motion of an isolated rod consists only of  small angular deflections about a fixed direction.  An isolated rod in a sea of sheared disks ceases to rotate, except at very low packings. 

We believe that the dependence of $-\langle\dot\theta_i\rangle/\dot\gamma$ and $S_2$ on the number of rods $N_\mathrm{rod}$ is closely related to the depletion forces that cause the rods to form parallel oriented clusters.  For a rod of length $\ell=(1+\alpha)D_b$ in a dense packing to rotate, it is necessary to have a local packing fluctuation on the length scale $\ell$, so that  sufficient free volume opens up to allow the rod to rotate.  Rods that are in parallel side-to-side contact have more local free volume than rods in isolation; that is the origin of the depletion force.  The sliding of one rod over another is a relatively low energy fluctuation that facilitates packing fluctuations on the length scale $\ell$, and so facilitates rod rotation.  In contrast, a rod in isolation from other rods is surrounded by disks; the motion of any one disk creates a packing fluctuation on the length scale $D$, and it would thus take a correlated motion of several disks to create sufficient free volume to allow the rod to rotate.  Such correlated spatial motion is rare, and consequently we find that for a system with only a single isolated rod, the rod ceases to rotate on the strain scale $\gamma\approx 100$ of our simulations.
But as the fraction of rods $N_\mathrm{rod}/N$ increases, the clustering of rods as measured by $Z_\mathrm{side}/Z_\mathrm{rod}$ increases (see Fig.~\ref{p-Z-S2-v-g}), and hence the rate of rotation, $-\langle\dot\theta_i\rangle/\dot\gamma$, increases.  The increasing rate of rotation then leads to a decrease in the magnitude of the nematic ordering $S_2$ \cite{MT2}.

In Fig.~\ref{rods-disks}(c) we plot the steady-state ensemble averaged value of the orientational angle $\theta_2$ of the nematic order parameter.  It is interesting that, in the dense region near jamming, as $N_\mathrm{rod}$ decreases the orientation angle $\theta_2$ increases, indicating a closer alignment of the rod with the direction of minimal stress, $\theta_-=45^\circ$.  Finally, in Fig.~\ref{rods-disks}(d) we show the steady-state average value for the rod clustering parameter $Z_\mathrm{side}/Z_\mathrm{rod}$ vs $\phi$, for systems with $N_\mathrm{rod}=64$, 128, 256 and 512 rods.  We see that $Z_\mathrm{side}/Z_\mathrm{rod}$, for fixed $N_\mathrm{rod}$, varies relatively little over the entire range of $\phi$.  

\section{Results: Pure Shear}

\begin{figure}
\centering
\includegraphics[width=3.3in]{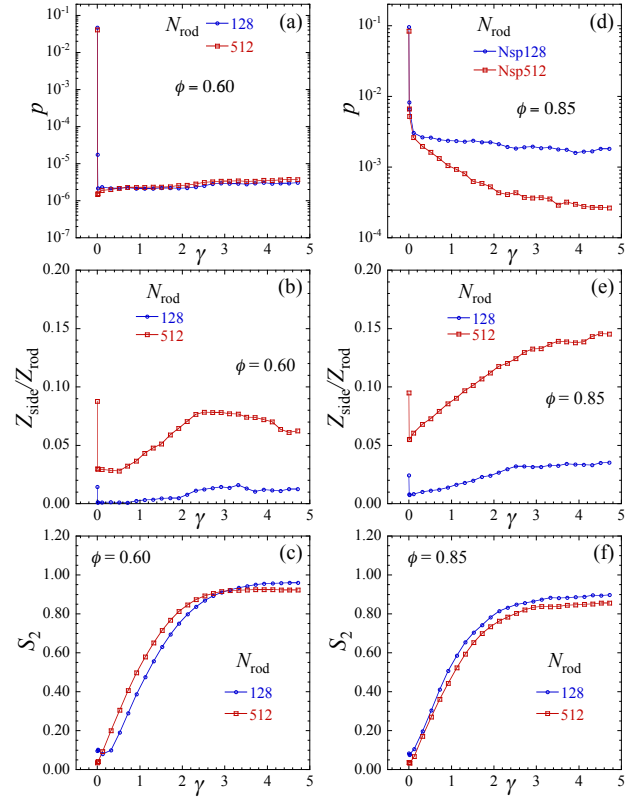}
\caption{For systems of $N_\mathrm{rod}=128$ and 512  rods of $\alpha=4$ in a sea of size-bidisperse circular disks with $N=2048$ total particles, undergoing pure shearing at a strain rate $\dot\gamma=10^{-5}$: (a) pressure $p$, (c) fraction of contacts on a spherocylinder that are side-to-side with another spherocylinder, $Z_\mathrm{side}/Z_\mathrm{rod}$, and (e) magnitude of the nematic order parameter $S_2$ vs net strain $\gamma=\dot\gamma t$, for packing $\phi=0.60$.  Similarly, (b) pressure $p$, (d) fraction $Z_\mathrm{side}/Z_\mathrm{rod}$, and (f) $S_2$ vs $\gamma$ for $\phi=0.85$.  Each data point, except for the few smallest, represents an average of the instantaneous values over a strain window of $\Delta\gamma=0.2$, so as to reduce fluctuations.
}
\label{pure-shear1} 
\end{figure}

Here we present our results for  the mixture of rods and disks in the case (ii) of pure shearing, defined by Eq.~(\ref{epureshear}).   The system is compressed in the $\mathbf{\hat y}$ direction, while expanded at the same rate in the $\mathbf{\hat x}$ direction, so that the system area remains constant.
As we have shown earlier in Ref.~\cite{MT2}, under pure shearing the orientation of rod shaped particles  relaxes to the direction of minimal stress $\theta_2=0$; there is no  continuous rotation of particles as occurs under simple shearing.  

Unlike simple shearing, where the Lees-Edwards boundary conditions allow us to shear to arbitrarily large total strains $\gamma$, in pure shearing one compresses in one direction  (here the $\mathbf{\hat y}$ direction) so the system will shrink to 
too narrow a height if one strains to too large  $\gamma$.  It is thus not always possible to shear long enough to reach the steady-state with finite system sizes \cite{MT2}.  We consider here mixtures with $N_\mathrm{rod}=128$ and 512 rods and $N=2048$ total particles.  To allow for a larger total strain, we start with a system of aspect ratio $L_y/L_x=12$, and shear until we reach $L_x/L_y=12$.  This allows us to reach a maximum total strain of $\gamma_\mathrm{max}=2\ln 12\approx 4.97$.  Our results  are for a shear rate $\dot\gamma=10^{-5}$ and are averaged over four independent runs starting from four different random initial configurations.

In Figs.~\ref{pure-shear1}(a) and \ref{pure-shear1}(b) we plot the pressure $p$ vs the net shear strain $\gamma=\dot\gamma t$ at the packings $\phi=0.60$, well below jamming, and at $\phi=0.85$, slightly above jamming.  Except for the few smallest $\gamma$ points, the data points here (and similarly for the other panels of Fig.~\ref{pure-shear1}) represent an average of the instantaneous values over a strain window of $\Delta\gamma=0.2$.   Since we can only shear to the relatively small $\gamma_\mathrm{max}\approx 5$, we see that our systems have not quite reached the steady state; the pressure $p$ continues to change gradually, rather than plateauing to a constant, at the largest $\gamma_\mathrm{max}$.
In Figs.~\ref{pure-shear1}(c) and \ref{pure-shear1}(d) we plot the rod clustering parameter $Z_\mathrm{side}/Z_\mathrm{rod}$ vs $\gamma$ for $\phi=0.60$ and 0.85, respectively.  Although we have not quite reached the steady state,  as the system is strained we see that $Z_\mathrm{side}/Z_\mathrm{rod}$  clearly increases   from the small value obtained immediately after the quench from the random initial configuration, thus indicating the presence of depletion forces.  As for the simple shearing shown in Fig.~\ref{p-Z-S2-v-g}, we see that $Z_\mathrm{side}/Z_\mathrm{rod}$ increases as $N_\mathrm{rod}$ increases, though for the smaller $\phi=0.60$ the values of $Z_\mathrm{side}/Z_\mathrm{rod}$ seem smaller than those found for simple shearing.  In Figs.~\ref{pure-shear1}(e) and \ref{pure-shear1}(f) we plot the magnitude of the nematic order parameter $S_2$ vs $\gamma$ for $\phi=0.60$ and 0.85, respectively.  As for the simple shearing in Fig.~\ref{p-Z-S2-v-g}, we see that the pure shearing orients the rod, causing $S_2$ to grow and saturate as the system approaches the steady-state.  

\begin{figure}
\centering
\includegraphics[width=3.3in]{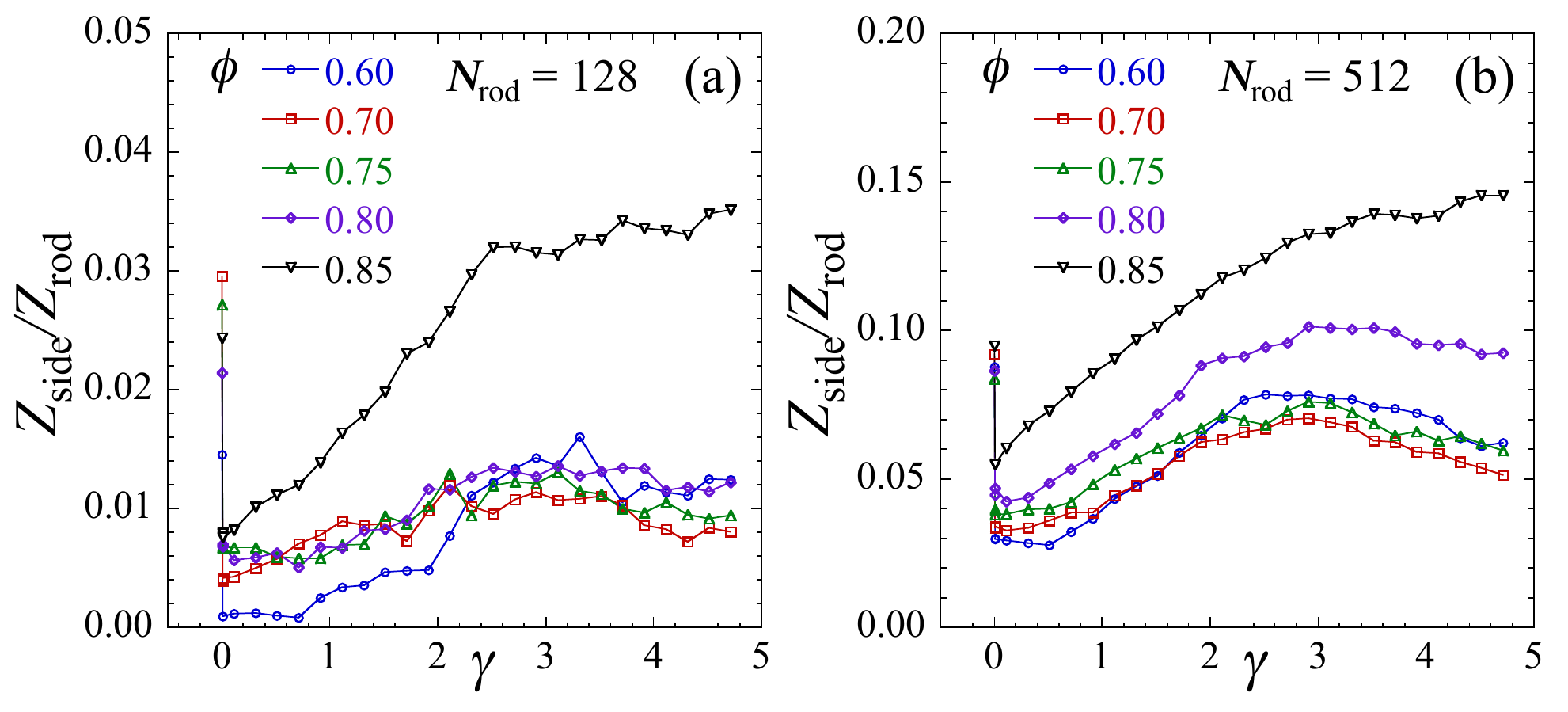}
\caption{For systems of $N_\mathrm{rod}$  spherocylindrical rods  of $\alpha=4$ in a sea of size-bidisperse circular disks with $N=2048$ total particles, undergoing pure shearing at a strain rate $\dot\gamma=10^{-5}$: fraction of contacts on a spherocylinder that are side-to-side with another spherocylinder, $Z_\mathrm{side}/Z_\mathrm{rod}$, vs net strain $\gamma=\dot\gamma t$, at several different packings $\phi=0.60$ to 0.85, for (a) $N_\mathrm{rod}=128$ and (b) $N_\mathrm{rod}=512$.  Each data point, except for the few smallest, represents an average of the instantaneous values over a strain window of $\Delta\gamma=0.2$, so as to reduce fluctuations.  
}
\label{pure-shear2} 
\end{figure}

\begin{figure}
\centering
\includegraphics[width=3.3in]{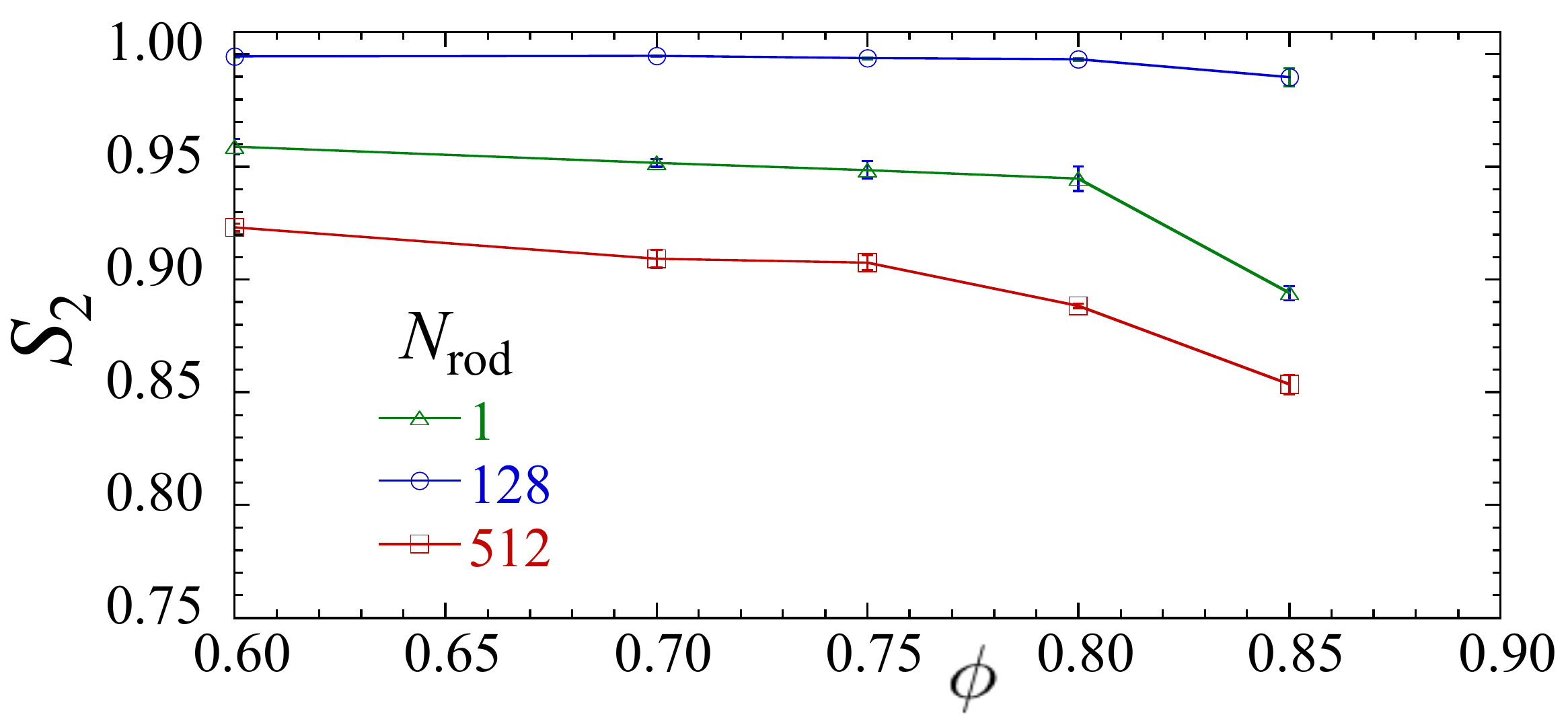}
\caption{For systems of $N_\mathrm{rod}=1$, 128, and 512  spherocylindrical rods  of $\alpha=4$ in a sea of size-bidisperse circular disks with $N=2048$ total particles, undergoing pure shearing at a strain rate $\dot\gamma=10^{-5}$:  magnitude of the nematic order parameter $S_2$ averaged over the range $4<\gamma<5$ at the end of the run, approximating the value in the steady-state, vs packing $\phi$.  As the value of $N_\mathrm{rod}$ increases, the curves go from top to bottom.
}
\label{s2-v-g-pureshear-ave} 
\end{figure}

In Fig.~\ref{pure-shear2}(a) and \ref{pure-shear2}(b) we plot the clustering parameter $Z_\mathrm{side}/Z_\mathrm{rod}$ vs $\gamma$ for $N_\mathrm{rod}=128$ and 512, respectively.  Here we show results for a range of different packings $\phi$.  Comparing the large $\gamma$ values of $Z_\mathrm{side}/Z_\mathrm{rod}$ seen here with the steady state values found in simple shear, shown in Fig.~\ref{rods-disks}(d), it seems that $Z_\mathrm{side}/Z_\mathrm{rod}$ varies more with the packing $\phi$ in pure shear as compared to simple shear.

Finally, although we have not quite reached the steady-state, the plots of $S_2$ in Figs.~\ref{pure-shear1}(e) and \ref{pure-shear1}(f) suggest that $S_2$ at the largest $\gamma$ is not far from its steady-state value.  For a rough estimate of that steady-state value we therefore compute as follows.  We first compute the ensemble average of $S_2$ for each individual run, averaging only over configurations in the strain window $4<\gamma<5$, at the end of the run.  We then average the resulting values of $S_2$ over the four different independent runs (for $N_\mathrm{rod}=1$ we use eight independent runs), and we estimate the statistical error from the variance of those values.  The resulting $S_2$ is plotted vs $\phi$ in Fig.~\ref{s2-v-g-pureshear-ave}.  We show results for $N_\mathrm{rod}=1$, 128, and 512.  As was found for simple shear in Fig.~\ref{rods-disks}(b), we find that $S_2$ decreases as $N_\mathrm{rod}$ increases.  We thus conclude that depletion forces are present in a pure sheared system, though at some packings they may be smaller than we have found in simple shearing.  Animations of pure shearing with different $N_\mathrm{rod}$ are available as additional Supplemental Material \cite{SM}.

\section{Results: Isotropic Compression}

\begin{figure}
\centering
\includegraphics[width=3.3in]{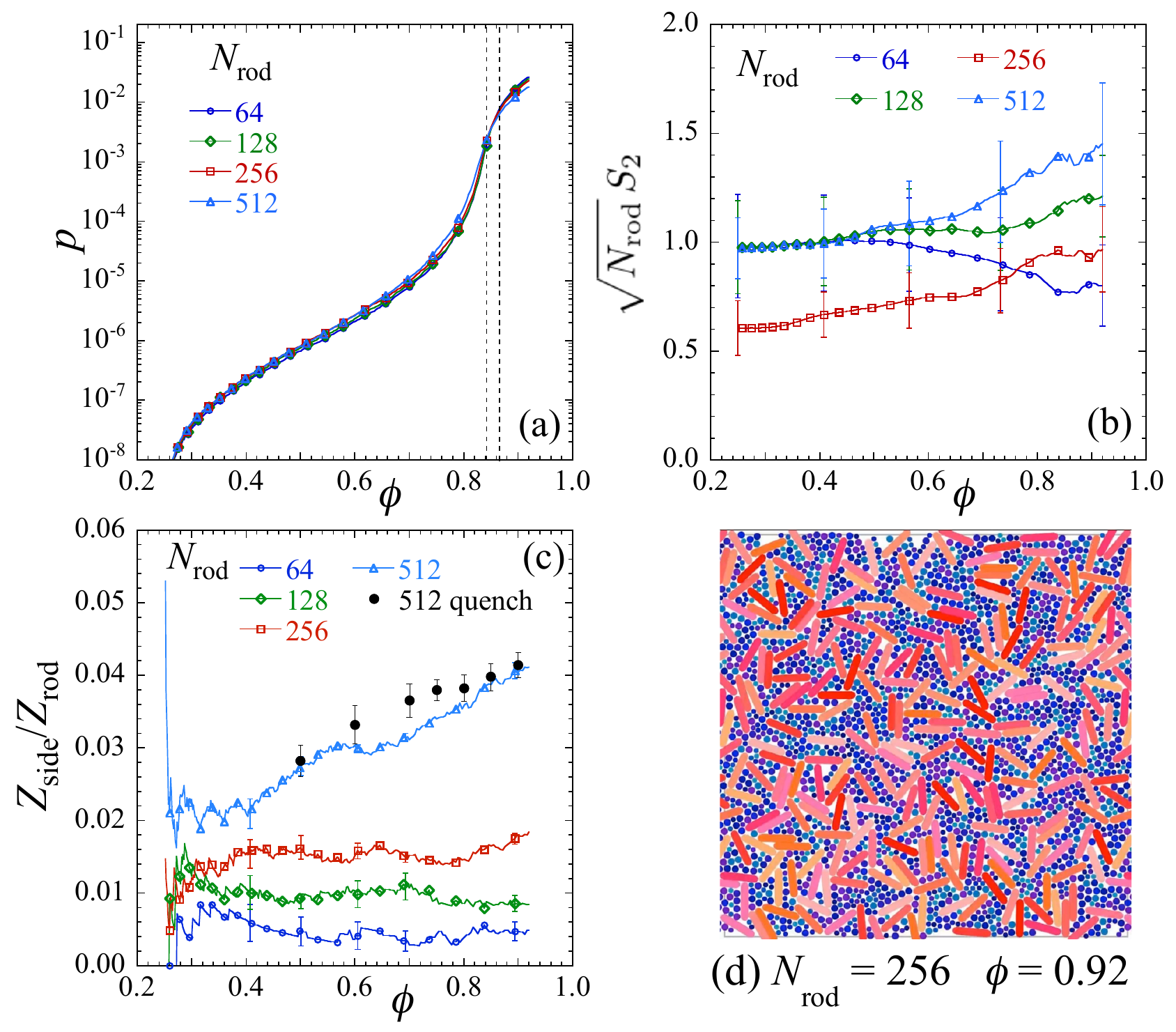}
\caption{For systems of $N_\mathrm{rod}=64$, 128, 256, and 512 spherocylindrical rods  of $\alpha=4$ in a sea of size-bidisperse circular disks with $N=2048$ total particles, sheared at a strain rate of $\dot\gamma=10^{-6}$: (a) pressure $p$, (b) magnitude of the nematic order parameter $S_2$ scaled by $\sqrt{N_\mathrm{rod}}$, and (c) rod clustering parameter $Z_\mathrm{side}/Z_\mathrm{rod}$ vs packing $\phi$. In (a) the vertical dashed lines indicate the jamming packings of  systems of only bidisperse circular disks, $\phi_J^{(0)}=0.8417$, and only bidisperse $\alpha=4$ spherocylinders, $\phi_J^{(4)}\approx0.866$.
In (c) solid black circles represent values obtained from relaxing a random configuration with $N_\mathrm{rod}=512$ at each $\phi$ without compression.  For clarity, in each panel symbols are shown only on a subset of data points.  In (a) error bars are typical smaller than the data point symbol; in (b) and (c) respresentative error bars are shown on a subset of the data points.  (d) Snapshot of a configuration of $N_\mathrm{rod}=256$ rods at the densest packing $\phi=0.92$.  In (c) the value of $N_\mathrm{rod}$ increases as the curves go from bottom to top.
}
\label{compress} 
\end{figure}

Finally we consider the behavior of the mixture of rods and disks in the case (iii) of isotropic compression, defined by Eq.~(\ref{ecompress}).  Our results here are for a compression rate of $\dot\gamma=10^{-6}$ and represent an average over eight independent runs starting from different random initial configurations.  At each compression step of strain increment $\Delta\gamma=\dot\gamma\Delta t$, the packing fraction increases by $\Delta \phi/\phi = 2\Delta\gamma$.  We will therefore plot our results vs $\phi$ rather than $\gamma$.  We start our compression runs from a random initial configuration at the dilute packing $\phi_\mathrm{init}=0.25$.

In Fig.~\ref{compress}(a) we plot the pressure $p$ vs $\phi$ for systems with the different values of $N_\mathrm{rod}$.  The vertical dashed lines indicate the compression-driven jamming packings of  systems of only bidisperse circular disks, $\phi_J^{(0)}=0.8417$ \cite{FSS}, and only bidisperse $\alpha=4$ spherocylinders \cite{MTCompress}, $\phi_J^{(4)}\approx0.866$.  Note that these values of $\phi_J$ for compression-driven jamming are lower than those for simple shear-driven jamming; this is particularly so for the case of spherocylinders, due to the nematic ordering that occurs for spherocylinders under shear \cite{MT2} but not under compression \cite{MTCompress}.
We see that, unlike the behavior of $p$ in simple shear, as shown in Fig.~\ref{p-eta-mu}(a), there is relatively little dependence of $p$ on $N_\mathrm{rod}$.  The small dependence that exists shows $p$ to increase as $N_\mathrm{rod}$ increases below $\phi_J^{(0)}$, but $p$ to decrease as $N_\mathrm{rod}$ increases above $\phi_J^{(0)}$.

The weak dependence of $p$ on $N_\mathrm{rod}$ we believe is due to the absence of orientational ordering of the compressed rods, as we have shown previously to be the case for a system of only size-bidisperse rods \cite{MTCompress}.  It is the ordering of the rods under shear that allows the system to pack more efficiently and to relax the pressure; this process is absent in compression.  We argue for the absence of orientational ordering of the rods as follows.  If the rods had completely random orientations, a finite number of rods would still possess some small finite nematic ordering as a statistical fluctuation.  However we would expect the magnitude of that nematic ordering to scale with the number of rods as $S_2\sim1/\sqrt{N_\mathrm{rod}}$, and so vanish in the infinite system limit.  In Fig.~\ref{compress}(b) we therefore plot $\sqrt{N_\mathrm{rod}}S_2$ vs $\phi$, for systems with $N_\mathrm{rod}=64$, 128, 256, and 512 rods.  Error bars are determined from the variance of values found in the eight independent compression runs.  We see that $\sqrt{N_\mathrm{rod}}S_2$ is, within the estimated errors, independent of $N_\mathrm{rod}$.  We thus conclude that the rods show no nematic ordering under compression.

Finally we consider the rod clustering parameter $Z_\mathrm{side}/Z_\mathrm{rod}$, which is plotted vs $\phi$ in Fig.~\ref{compress}(c).  For $N_\mathrm{rod}\le 256$ we see that $Z_\mathrm{side}/Z_\mathrm{rod}$ barely changes as the system is compressed and $\phi$ increases.  For $N_\mathrm{rod}=512$, however, we see a steady increase in $Z_\mathrm{side}/Z_\mathrm{rod}$ with increasing $\phi$, although the values of $Z_\mathrm{side}/Z_\mathrm{rod}$ found remain small compared to those found in shearing.  To determine if this increase in $Z_\mathrm{side}/Z_\mathrm{rod}$ is due to the development of depletion forces as the system is compressed, or whether it is just an effect of the increasing density of particles, we do the following.  At each value of $\phi=0.5$, 0.6, \dots, 0.90 we create a random initial configuration with $N_\mathrm{rod}=512$ rods in the same manner that we do for simple shearing.  We then relax the energy of that configuration by simulating  the equations of motion Eqs.~(\ref{er}) and (\ref{etheta}), only setting $\mathbf{v}_\mathrm{host}=0$ so there is no compression.  This relaxation reduces the unphysically large particle overlaps of the initial random configuration, spreading the particles more evenly throughout the system, but without inducing any correlations that might be created by compression.  The values of $Z_\mathrm{side}/Z_\mathrm{rod}$ so obtained are shown as the solid black circles in Fig.~\ref{compress}(c).  We see that these values roughly approximate (indeed they are slightly larger than) the values obtained by compression of the initial dilute configuration.  
We thus conclude that the increasing $Z_\mathrm{side}/Z_\mathrm{rod}$ found for $N_\mathrm{rod}=512$ is simply an effect of the increasing density of particles.  
Indeed, the values of $Z_\mathrm{side}/Z_\mathrm{rod}$ found here for compression are comparable to the values found in Fig.~\ref{p-Z-S2-v-g} for simple shearing, if one looks just after the  rapid quench of the large overlaps in the initial random state, but before the increase in $Z_\mathrm{side}/Z_\mathrm{rod}$ that results from the shearing.
We conclude that no depletion forces develop from athermal isotropic compression of mixtures of rods and disks.  Finally, in Fig.~\ref{compress}(d) we show a snapshot of a configuration of $N_\mathrm{rod}=256$ rods, compressed above jamming to the packing $\phi=0.92$.  Visual inspection is consistent with our result that there is no nematic ordering of the rods, and little tendency for them to group into parallel clusters.  Animations of compressions with different $N_\mathrm{rod}$ are available as additional Supplemental Material \cite{SM}.

\section{Discussion}

We have presented results for the behavior of athermal mixtures of frictionlesss circular disks and moderately elongated rods in two dimensions, undergoing three different types of linear elastic deformations at a fixed small strain rate: (i) simple shearing, (ii) pure shearing, and (iii) isotropic compression.  We have looked for evidence for depletion forces acting between the rods, as measured by the number of side-to-side contacts between rods that develop as the system approaches steady-state.  We find that such depletion forces do appear under both simple and pure shearing, but not under isotropic compression.  

For simple shearing we have explicitly shown that the side-to-side contacts are not  simply a  manifestation of the nematic ordering that the rods undergo when sheared, but rather they depend directly on the presence of the disks in which the rods are immersed.  As the density of disks increases at fixed rod packing $\phi_\mathrm{rod}$, the number of  side-to-side contacts between the rods increases (see Fig.~\ref{Zside}). For simple shearing we have also shown the following.  For systems held at constant pressure, the  viscosity of the mixture decreases slightly as the fraction of rods increases (see Fig.~\ref{p-eta-mu}(d)).  As the fraction of rods decreases, the average angular velocity of the rods decreases, while the magnitude of the nematic ordering increases (see Fig.~\ref{rods-disks}).  A single isolated rod in a sea of disks ceases to rotate at all, except at very low packings.

In a recent experimental work \cite{hopper}, it was observed that the addition of elongated rod shaped particles to a quasi-2D granular system of glass beads increased the rate of discharge of the beads in hopper flow.  As the number of rods initially increased, the rate of discharge increased.   It was argued that the mechanism for this increasing discharge rate is the rotation of the rods near the surface layer that causes a secondary flow of the glass beads and a significant increase in the thickness of the flowing layer.  While our simulations are spatially uniform and have no surface layer, our observation that increasing the fraction of rods increases the clustering of rods, which then results in a decrease in the shear viscosity at constant pressure (see Fig.~\ref{p-eta-mu}(d)) as well as an increase in the  average rate of rod rotation, may play some role in this effect.

\begin{acknowledgments}
We thank Karen Daniels and Peter Olsson for helpful discussion.  This work was supported by National Science Foundation Grant No. CBET-1435861 and No. DMR-1809318. Computations were carried out at the Center for Integrated Research Computing at the University of Rochester. 
\end{acknowledgments}

\bibliographystyle{apsrev4-1}

\end{document}